\documentclass[12pt]{article}
\title{On the coefficients of the B\'{a}ez-Duarte criterion for the Riemann hypothesis and their extensions} %TBR
\author{Mark W. Coffey\\
Department of Physics\\
Colorado School of Mines\\
Golden, CO  80401\\
\{mcoffey@mines.edu\} \\
(Received $\mbox{~~~~~~~~~~~~~~~~~~~~~~~~~~~~~~~2006}$)}
\date{September 6, 2006}
\begin{document}
\maketitle
%\vspace{.25cm}
\baselineskip=25 pt
\begin{abstract}

We present analytic properties and extensions of the constants $c_k$ appearing in
the B\'{a}ez-Duarte criterion for the Riemann hypothesis.  These constants are the coefficients of Pochhammer polynomials in a series representation of the reciprocal
of the Riemann zeta function.  We present generalizations of this representation to the Hurwitz zeta and many other special functions. We relate the corresponding coefficients to other known constants including the Stieltjes constants and present summatory relations.  In addition, we generalize the Ma\'{s}lanka hypergeometric-like representation for the zeta function in several ways.    

\end{abstract}
 
\vspace{.25cm}
\baselineskip=15pt
\centerline{\bf Key words and phrases}
\medskip 

\noindent
B\'{a}ez-Duarte criterion, Riemann hypothesis, Stieltjes constants,
Pochhammer polynomial, Riemann and Hurwitz zeta functions, polygamma function,
complete Bell polynomials, Ma\'{s}lanka representation, Dirichlet $L$ function

%\vspace{.25cm}
%\vfill
%\centerline{\bf AMS Classification Numbers}
%11M26, 11M35     %RH, Hurwitz and Lerch zeta funcs resp. 
 
\baselineskip=25pt
\pagebreak
\medskip
\centerline{\bf Introduction}
\medskip
%
%consider using a framework of approximately 1/B_n(x) or 1/B_{2n}(x) as
%summand in an alt. binomial sum--can work out something relative to the
%Baez-Duarte criterion for the RH ??
%
As reformulated by B\'{a}ez-Duarte \cite{baez1}, the Riemann hypothesis (RH) is 
equivalent to a certain growth condition on constants $c_k$ that appear in a
series representation of the reciprocal of the Riemann zeta function $\zeta(s)$.
In this paper we analytically investigate these constants and introduce some
parametrized extensions.  We point out the importance of such extensions for 
future work.  We relate parametrized coefficients $c_k(b,a)$ to other important
constants of analytic number theory including the Stieltjes constants and
present summatory relations for the former.

With 
$$c_k \equiv \sum_{j=0}^k (-1)^j {k \choose j} {1 \over {\zeta(2j+2)}}, ~~~~~~
k \geq 0, \eqno(1)$$
unconditionally $c_k = O(k^{-1/2})$ and on the RH we have $c_k = O(k^{-3/4
+\epsilon})$ for any $\epsilon >0$ \cite{baez1}.  The constants $c_k$ are known to consist of a relatively rapidly decreasing term $\propto -1/k^2$ and an
oscillatory contribution, as can be shown by Rice's integrals or other asymptotic
methods \cite{maslanka,wolf}.  Not surprisingly, it is the detailed behaviour
of the oscillatory contribution upon which the validity of the RH depends. 
Numerical results for $c_k$ are presented in Refs. \cite{beltra,cislo,maslanka,wolf} 
and the first billion values have been reported.  These values are consistent
with the B\'{a}ez-Duarte criterion under the RH.  If further $c_k=O(k^{-3/4})$, then
the complex zeros of $\zeta(s)$ are on the critical line Re $s=1/2$ and are
simple.

\medskip
\centerline{\bf Summatory relations for $c_k$}
\medskip

In this section we relate $c_k$ to the Stieltjes constants $\gamma_k$.  The
latter are the coefficients of the Laurent series of the Riemann zeta function
about $s=1$.  In preparation we have Lemma 1 concerning the derivatives of
Pochhammer polynomials.

Let $P_k(s) \equiv (1-s)_k/k!= (-1)^k\Gamma(s)/k!\Gamma(s-k)$, 
where $(a)_n = \Gamma(a+n)/\Gamma(a)$ is the Pochhamer symbol.  Let 
$\psi=\Gamma'/\Gamma$ be the digamma function, where $\Gamma$ is the Gamma
function, and $\psi^{(j)}$ the polygamma function \cite{andrews}.

{\bf Lemma 1}.  Set
$$g(s) \equiv {1 \over 2}\left[\psi(1-s/2)-\psi(k+1-s/2)\right], \eqno(2a)$$
with 
$$g^{(\ell)}(s) = {{(-1)^\ell} \over 2^{\ell+1}}\left[\psi^{(\ell)}(1-s/2)
-\psi^{(\ell)}(k+1-s/2)\right], \eqno(2b)$$		
Then we have
$${d \over {ds}}P_k\left({s \over 2}\right)=P_k\left({s \over 2}\right) g(s), \eqno(3)$$
and
$$\left({d \over {ds}}\right)^jP_k\left({s \over 2}\right)=P_k\left({s \over 2}\right) Y_j\left[g(s),g'(s),\ldots,g^{(j-1)} (s)\right], \eqno(4)$$
where $Y_j$ are (exponential) complete Bell polynomials \cite{comtet}.

{\em Proof}.  Equation 3 follows from the definition of the Pochhammer symbol
and Eq. (4) by Lemma 1 of Ref. \cite{coffeyharm06}.  

A very small subset of the relations between the constants $c_k$ and the Stieltjes
constants is contained in the following.  Equation (5c) presents how the Euler
constant $\gamma=-\psi(1)$ may be written in terms of $c_j$.
{\newline \bf Proposition 1}.  We have
$$\sum_{k=0}^\infty {{\Gamma(k+1/2)} \over {k!}}c_k=0, \eqno(5a)$$
$$-{1 \over {2\sqrt{\pi}}}\sum_{k=0}^\infty {{\Gamma(k+1/2)} \over {k!}}
\psi(k+1/2)c_k=1, \eqno(5b)$$
and
$$\sum_{k=0}^\infty {{\Gamma(k+1/2)} \over {k!}}[\psi^2(k+1/2)+\psi'(k+1/2)]c_k
=-4\sqrt{\pi}(\gamma-2\ln 2).  \eqno(5c)$$

As prelude to the proof of Proposition 1, we know that the representation
$${1 \over {\zeta(s)}} = \sum_{k=0}^\infty c_k P_k(s/2), \eqno(6)$$
holds unconditionally in the half plane Re $s > 1$, converging uniformly
on compact sets \cite{baez1}.  
%We assume for Proposition 1 something stronger.  Namely, 
%{\newline \bf Conjecture 1}. The series representation (6) holds in a neighborhood
%of the point $s=1$.

%In fact we believe that it will be possible to show the following
%{\newline \bf Conjecture 1}.  The series representation (6) holds unconditionally
%in the disc $|s-1| < 1/2$.

%Of course on the RH, Conjecture 1 follows.  However, within the stated disc 
%$\zeta(s)$ is well known to be zero free and a separate demonstration may be
%possible.

{\em Proof of Proposition 1}.  We combine the definition of the Stieltjes
constants (e.g., \cite{ivic}) with Eq. (6) to write
$${1 \over {\zeta(s)}}=\left[{1 \over {s-1}}+\sum_{k=0}^\infty {\gamma_k \over {k!}}
(s-1)^k \right ]^{-1}=\sum_{k=0}^\infty c_k P_k(s/2)  \eqno(7a)$$
$$=s-1-\gamma(s-1)^2+(\gamma^2+\gamma_1)(s-1)^3+(-\gamma^3-2\gamma\gamma_1-\gamma_2/2)
(s-1)^4 + O[(s-1)^5].  \eqno(7b)$$
Here we used $\gamma_0=\gamma$.  We then Taylor expand the right side of Eq. (7a)
about $s=1$ using Lemma 1.  In particular, we have
$$\left.\left({d \over {ds}}\right)^j P_k\left({s \over 2}\right)\right|_{s=1} ={{\Gamma(k+1/2)} \over {\sqrt{\pi}k!}}Y_j\left[g(s),g'(s),\ldots,g^{(j-1)}
(s)\right]_{s=1}, \eqno(8)$$
$$g(1) = {1 \over 2}\left[\psi(1/2)-\psi(k+1/2)\right]=-{{\Gamma(k+1/2)} \over {2\sqrt{\pi}k!}}\sum_{\ell=0}^{k-1} {1 \over {\ell+1/2}}, \eqno(9a)$$
and
$$g^{(\ell)}(1) = {{(-1)^\ell} \over 2^{\ell+1}}\left[\psi^{(\ell)}(1/2)
-\psi^{(\ell)}(k+1/2)\right]=-{{\ell !} \over {2^{\ell+1}}}\sum_{j=0}^{k-1}
{1 \over {(j+1/2)^{\ell+1}}}. \eqno(9b)$$
Equations (9a) and (9b) follow from the functional equations of the digamma and
polygamma functions respectively.  We next equate successive like powers of $s-1$
on both sides of Eq. (7).  Effectively, we evaluate the successive derivatives of
the representation (6) as $s \to 1^+$.  We first obtain Eq. (5a).  
We then use it to obtain Eq. (5b).  We then use both of these to obtain Eq. (5c)
and Proposition 1 follows.  

Remarks.  The continuation of the process just described yields the explicit relation between $\gamma_k$ and sums over the constants $c_j$.  The appearance of the low order Bell polynomials $Y_0=1$, $Y_1(x_1)=x_1$, and $Y_2(x_1,x_2)=x_1^2+x_2$ is implicit in writing Eqs. (5a)-(5c).  They do not appear in the final results there since we performed successive manipulations.  

Equation (5a) is a reflection of the simple pole of $\zeta(s)$ at $s=1$ and of the relation $\sum_{n=1}^\infty \mu(n)/n=0$, where $\mu$ is the M\"{o}bius function.  Once results such as Eqs. (5) have been derived, they may be directly verified using the alternative expression \cite{baez1} 
$$c_k=\sum_{n=1}^\infty {{\mu(n)} \over n^2}\left(1-{1 \over n^2}\right)^k.  \eqno(10)$$
For instance, we recover
$$\sum_{k=0}^\infty {{\Gamma(k+1/2)} \over {k!}} c_k = \sqrt{\pi} \sum_{n=1}^\infty
{{\mu(n)} \over n^2} n = 0.  \eqno(11)$$

{\bf Corollary 1}.  In Eq. (5a), the ratio $\Gamma(k+1/2)/k! \sim k^{-1/2}$ for
$k \to \infty$ while in Eq. (5b) the factor $\psi(k+1/2) \sim \ln k$ as $k \to \infty$.  Therefore, the former equation shows that if the $c_k$'s did not change
sign, they would have to decrease at least as fast as $k^{-1/2}$ as $k \to \infty$.

The following expression may be combined with the Stieltjes constant expansion of
$1/\zeta(s)$ given in Eq. (7) to write $\gamma_j$ in terms of sums of the constants $c_k$ and Stirling numbers of the first kind $s(k,\ell)$ \cite{comtet,riordan1,riordan2}.
{\newline \bf Lemma 2}.  The coefficient of $(s-1)^j$ on the right side of Eq. (7a) is
given by
$$\sum_{k=0}^j (-1)^k {c_k \over {k!}} \sum_{\ell=j}^k {{s(k,\ell)} \over 2^\ell}
(-1)^{\ell-j} {\ell \choose j}.  \eqno(12)$$

{\em Proof}.  We first re-express the Pochhammer polynomials using the Stirling
numbers $s(k,\ell)$, then binomially expand and reorder sums:
$$\sum_{k=0}^\infty {c_k \over {k!}} (1-s/2)_k
=\sum_{k=0}^\infty {c_k \over {k!}} \sum_{\ell=0}^k (-1)^{k+\ell}s(k,\ell)(1-s/2)^\ell$$
$$=\sum_{k=0}^\infty (-1)^k {c_k \over {k!}}\sum_{\ell=0}^k {{s(k,\ell)} \over 2^\ell}
\sum_{j=0}^\ell (-1)^{\ell-j} {\ell \choose j} (s-1)^j$$
$$=\sum_{k=0}^\infty (-1)^k {c_k \over {k!}}\sum_{j=0}^k \sum_{\ell=j}^k {{s(k,\ell)} \over 2^\ell} (-1)^{\ell-j} {\ell \choose j} (s-1)^j.  \eqno(13)$$
  
\medskip
%\pagebreak
\centerline{\bf First extension of the constants $c_k$}
\medskip

The Hurwitz zeta function $\zeta(s,a) = \sum_{k=0}^\infty (k+a)^{-s}$, $a \notin N_0^-$ for Re $s>1$ extends to an analytic function with only a simple pole at $s=1$.  Here we demonstrate the representation
{\newline \bf Corollary 2}.  For Re $s>1$ we have
$${1 \over {\zeta(s,a)}} = \sum_{k=0}^\infty c_k(a) P_k(s/2), \eqno(14)$$
with
$$c_k(a) = \sum_{j=0}^k (-1)^j {k \choose j} {1 \over {\zeta(2j+2,a)}}, ~~~~~~
k \geq 0. \eqno(15)$$
%The series representation (14) should hold for at least Re $s>1$ and on the RH
%for Re $s>1/2$.  
In particular, there results at $a=1/2$
{\newline \bf Corollary 3}.  
$${1 \over {\zeta(s)}}=(2^s-1)\sum_{k=0}^\infty c_k(1/2) P_k(s/2).  \eqno(16)$$

Corollary 2 follows from 
{\newline \bf Proposition 2}.  Define for $b > 1+\delta$ with $\delta >0$ and 
Re $a>0$ the functions
$$F(x,b,a) \equiv \sum_{k=1}^\infty {{(-1)^{k+1}} \over {\zeta(bk,a)}} {x^k \over
{(k-1)!}} \eqno(17)$$
and
$$\varphi(s,b,a) \equiv \int_0^\infty x^{-(s/b+1)} F(x,b,a)dx, ~~~~~~
1 < \mbox{Re} ~s < b.  \eqno(18)$$
Put
$$c_k(b,a) \equiv \sum_{j=0}^k (-1)^j {k \choose j} {1 \over {\zeta(bj+b,a)}}.
\eqno(19)$$
Then we have the series and integral representations for Re $s>1$
$${1 \over {\zeta(s,a)}} = \sum_{k=0}^\infty c_k(b,a) P_k(s/b) = {{\varphi(s,b,a)}
\over {\Gamma(1-s/b)}}.  \eqno(20)$$

In the proof we use the infinite series
{\newline \bf Lemma 3}
$$\sum_{k=j}^\infty {1 \over {k!}}{k \choose j} x^k=\sum_{k=0}^\infty {1 \over {(k+j)!}} {{k+j} \choose j} x^{k+j} = {x^j \over {j!}} e^x,  \eqno(21)$$
and
{\newline \bf Lemma 4}
$$\sum_{n=0}^\infty F\left[{x \over {(n+a)^b}},b,a \right]=x e^{-x}.  \eqno(22)$$
The series (17) is uniformly convergent on compact sets of the complex $x$ plane 
so that the interchange of sums used to show Eq. (22) is valid.

We have
$$\int_0^\infty x^{-(s/b+1)} F\left[{x \over {(n+a)^b}},b,a \right ]dx
=(n+a)^{-s} \varphi(s,b,a).  \eqno(23)$$
Without assuming the RH, $\varphi$ defined in Eq. (18) converges absolutely and
uniformly in the strip specified.  The behaviour $F(x,b,a) \sim x/\zeta(bk,a)$
as $x \to 0$ dictates the condition Re $s < b$.  The behaviour $F(x,b,a)=o(x^{1/b})$
as $x \to \infty$ without the RH gives the requirement Re $s > 1$.
Then summing both sides of Eq. (23) on $n$ from $0$ to $\infty$ and using
Lemma 4 gives
$$\zeta(s,a) \varphi(s,b,a) = \int_0^\infty x^{-s/b} e^{-x} dx = \Gamma(1-s/b),
\eqno(24)$$
so that we have obtained the 'outer' equality of Eq. (20).

We next re-express the function $\varphi$ in terms of Pochhammer polynomials.
We have from Eqs. (17) and (18)
$$\varphi(s,b,a)=\int_0^\infty \sum_{j=0}^\infty {{(-1)^j} \over {\zeta(bj+b,a)}}
{1 \over {j!}} x^{j-s/b}dx \eqno(25a)$$
$$=\int_0^\infty \sum_{j=0}^\infty {{(-1)^j} \over {\zeta(bj+b,a)}}
{e^x \over {j!}} e^{-x} x^{j-s/b}dx. \eqno(25b)$$
We next apply Lemma 3 so that
$$\varphi(s,b,a)=\int_0^\infty \sum_{j=0}^\infty {{(-1)^j} \over {\zeta(bj+b,a)}}
\sum_{k=j}^\infty {1 \over {k!}} {k \choose j} e^{-x} x^{k-s/b} dx \eqno(26a)$$
$$=\int_0^\infty \sum_{k=0}^\infty {1 \over {k!}} \sum_{j=0}^k 
{{(-1)^j} \over {\zeta(bj+b,a)}} {k \choose j} e^{-x} x^{k-s/b} dx \eqno(26b)$$
$$=\sum_{k=0}^\infty {1 \over {k!}} c_k(b,a) \Gamma(k+1-s/b).  \eqno(26c)$$
In obtaining Eq. (26b) from (26a) we reordered the double series and (26c) used
the definition (19).  We have therefore found that
$$\varphi(s,b,a) = \Gamma(1-s/b) \sum_{k=0}^\infty c_k(b,a) P_k(s/b),  \eqno(27)$$ 
and Proposition 2 is completed.

%We could relax the condition that $b \geq 2$ be an integer in Proposition 2,
%and even more so to $b > 1 + \delta$ for $\delta > 0$, but this is unnecessary
%for our current purpose.    %have now done that ...
We have shown a way to directly relate the RH criteria of B\'{a}ez-Duarte \cite{baez1} 
and of Riesz \cite{riesz}.  The summatory function appearing in the Riesz
criterion corresponds to $R(x) \equiv F(x,2,1)$ in Eq. (17) and under the RH it
is $O(x^{1/4+\epsilon})$ for $\epsilon > 0$.

A recent construction for the function $1/\zeta$ similar to Proposition 2 has
been given in Ref. \cite{beltra}.  The authors of that reference used the
M\"{o}bius function in that development, whereas we have proceeded differently
and obtained a result also applying to the reciprocal of the Hurwitz zeta
function.

Arguing as we have in Proposition 2 gives many extensions.  An example is
{\newline \bf Corollary 4}.  Putting
$$G(x) \equiv \sum_{k=1}^\infty {{(-1)^k} \over {k!}} {x^k \over {\zeta(2k+1)}},
\eqno(28)$$
$$\varphi_G(x)=\int_0^\infty x^{-(s+1)/2} G(x)dx, \eqno(29)$$
and 
$$c_k^G \equiv \sum_{j=1}^k (-1)^j {k \choose j} {1 \over {\zeta(2j+1)}}, \eqno(30)$$
we have for Re $s > 1$ the representations
$${1 \over {\zeta(s)}} = \sum_{k=1}^\infty c_k^G P_k[(s+1)/2]
={{\varphi_G(s)} \over {\Gamma[(1-s)/2]}}.  \eqno(31)$$

Given Corollary 2 and Proposition 2 Eqs. (7) extend to
$${1 \over {\zeta(s,a)}}=\left[{1 \over {s-1}}+\sum_{k=0}^\infty {{\gamma_k(a)} \over {k!}} (s-1)^k \right ]^{-1}=\sum_{k=0}^\infty c_k(b,a) P_k(s/b)$$
$$=s-1+\psi(a)(s-1)^2+[\psi^2(a)+\gamma_1(a)](s-1)^3+[-\psi^3(a)+2\psi(a)\gamma_1(a)
-\gamma_2(a)/2](s-1)^4 + O[(s-1)^5],  \eqno(32)$$
wherein $\gamma_0(a)=-\psi(a)$ has been used.  The Stieltjes constants $\gamma_k(a)$
may be written in terms of sums containing the Bernoulli numbers $B_j$ and  
elementary constants such as powers of $\ln 2$.  Further properties of $\gamma_k(a)$
are given in the very recent Refs. \cite{coffeykremin} and \cite{coffeyprsa}.

Remark 1.  Analogous to Eqs. (14) and (15) it is not difficult to show that the
Ma\'{s}lanka representation for $\zeta(s)$ \cite{maslanka} may be extended to
$$\zeta(s,a)={1 \over {s-1}}\sum_{k=0}^\infty (1-s/2)_k {{A_k(a)} \over {k!}},
\eqno(33)$$
where
$$A_k(a) \equiv \sum_{j=0}^k (-1)^j {k \choose j}(2j+1)\zeta(2j+2,a).  \eqno(34)$$
As a Corollary, we obtain as special cases a representation for Bernoulli
polynomials $B_n(x) = -n \zeta(1-n,x)$ and polygamma functions
$\psi^{(m)}(z)=(-1)^{m+1}m!\zeta(m+1,z)$.  Additional cases include 
representations for the alternating Hurwitz zeta function, the digamma function,
and the function 
$\beta(x)=(1/2)[\psi[(x+1)/2]-\psi(x/2)]=\sum_{k=0}^\infty (-1)^k/(x+k)$.
%later ... maybe make as a display equation and put an equ. no. on it ...

Remark 2.  We believe that it is very useful to have the constants $c_k$
extended to include one or more parameters.  In that case manipulations on $c_k(a)$
and $1/\zeta(s,a)$ for instance may be performed with respect to $a$ and/or $s$
and then the Riemann zeta function case recovered by putting $a=1$ or $1/2$.
For instance, we have from Eq. (15) 
$${d \over {da}}c_k(a) = \sum_{j=0}^k (-1)^j {k \choose j} {{(2j+2)\zeta(2j+3,a)} 
\over {\zeta^2(2j+2,a)}}.  \eqno(35)$$

Remark 3.  The representation (33) may be further extended to the Hurwitz-Lerch
zeta function $\Phi(z,s,a)=\sum_{n=0}^\infty z^n/(n+a)^s$,
where $s \in C$ for $|z| <1$, Re $s>1$ when $|z|=1$.  In this case we have
{\newline \bf Proposition 3}
$$\Phi(z,s,a)={1 \over {s-1}}\sum_{k=0}^\infty (1-s/2)_k {{A_k(z,a)} \over {k!}},
\eqno(36)$$
where
$$A_k(z,a) \equiv \sum_{j=0}^k (-1)^j {k \choose j}(2j+1)\Phi(z,2j+2,a).  
\eqno(37)$$
The proof of Proposition 3 and an extension of it is contained as a special
case of Proposition 4 proved below.  As a Corollary, we obtain Ma\'{s}lanka-type
representations for polylogarithm functions (Jonqui\`{e}re's function)
$$\mbox{Li}_s(z)=z\Phi(z,s,1), \eqno(38)$$
where $s \in C$ for $|z| <1$, Re $s>1$ when $|z|=1$.

Remark 4.  We have
\newline{\bf Conjecture 1}.  For a class $\cal{M}$ of analytic functions 
expressible as a Dirichlet series and possessing at most polar singularities 
in the complex plane such that for $f \in {\cal{M}}$ and $q > 1$ we have the representation
$$f(s) = {1 \over {s-1}}\sum_{k=0}^\infty A_k P_k(s/q),
\eqno(39)$$
holding in a half plane of C with
$$A_k \equiv \sum_{j=0}^k (-1)^j {k \choose j}(qj+q-1)f(qj+q).  \eqno(40)$$
We believe that a demonstration of some form of this Conjecture is possible by
an appropriate application of sampling theory (cf. the Appendix).  However,
a more expedient approach may be to apply known results for approximating 
analytic functions in terms of the zeta function.  Then given the representation 
(33) Conjecture 1 would follow.
%give a list of other examples ???

%Should Conjecture 2 hold, there is probably an analogous generalization 
%of representation (14).

In support of Conjecture 1 we have the following
{\newline \bf Proposition 4}.  Let 
$$f(s,a)=\sum_{n=0}^\infty {f_n \over {(n+a)^s}}, \eqno(41)$$
where it is assumed that $\{f_n\}_0^\infty$ is such that the series converges
in a half-plane Re $s>\sigma>1$ and $a \in C/\{0,-1,\ldots\}$.  Then we have
for $p > 1$ 
$$f(s,a) = {1 \over {s-1}}\sum_{k=0}^\infty A_k(a) P_k(s/p), \eqno(42)$$
with
$$A_k(a) \equiv \sum_{j=0}^k (-1)^j {k \choose j}(pj+p-1)f(pj+p,a).  \eqno(43)$$

{\em Proof}.  We proceed as in Ref. \cite{baez2} in the case $f_n=1$, $a=1$,
and $p=2$, forming
$$\alpha^{-s} (s-1)f(s,a)=-{\partial \over {\partial \alpha}}\alpha^{1-s}
\sum_{n=0}^\infty {f_n \over {(n+a)^s}}. \eqno(44)$$
After term-by-term differentiation of the series we evaluate at $\alpha=1$.
The interchange of various operations is justified by the assumption on the
sequence $f_n$ and the estimates of Ref. \cite{baez2}.  

As a Corollary, we obtain generalized Ma\'{s}lanka-type representations for 
other important special functions.  These include the multiple zeta function
$$\zeta_n(s,z)=\sum_{k_1=0}^\infty \cdots \sum_{k_n=0}^\infty {1 \over {(k_1+
k_2 +\ldots+k_n+z)^s}}=\sum_{k=0}^\infty {1 \over {(k+z)^s}}{{k+n-1} \choose
{n-1}}, \eqno(45)$$
and the multiple Gamma function $\Gamma_n$.  The latter function has a product
representation and may be defined by the recurrence-functional equation $\Gamma_{n+1}(z+1)=\Gamma_{n+1}(z)/\Gamma_n(z)$, $\Gamma_1(z)=\Gamma(z)$,
$\Gamma_n(1)=1$, for $z \in C$ and $n \in N^+$.  The multiple Gamma function
may be expressed in terms of derivatives of the multiple zeta function
\cite{vardi}:
$$\ln \Gamma_n(z)=\lim_{s \to 0} {{\partial \zeta_n(s,z)} \over {\partial s}}
+\sum_{k=1}^n (-1)^k {z \choose {k-1}} R_{n+1-k}, \eqno(46a)$$
where 
$$R_n=\sum_{k=1}^n \lim_{s \to 0} {{\partial \zeta_n(s,1)} \over {\partial s}}.
\eqno(46b)$$

\medskip
%\pagebreak
\centerline{\bf Second extension of the constants $c_k$}
\medskip

Another possible extension of the constants $c_k$ would be to write
$$c_k(a,b) \equiv \sum_{j=0}^k (-1)^j {k \choose j} {1 \over {\zeta(aj+b)}}, ~~~~~~
k \geq 0, \eqno(47)$$
wherein we considered $a=b$ in Eq. (19).
This extension has also been observed in Ref. \cite{beltra} and numerical 
experiments presented.  However, we remain with the case $a=b=2$ and instead
note that
$$\zeta(2j+2)={{(2\pi)^{2j+2} (-1)^j B_{2j+2}} \over {2(2j+2)!}}=
{{(2\pi)^{2j+2} (-1)^j} \over {2(2j+2)!}}\left. B_{2j+2}(x)\right|_{x=0}.  
\eqno(48)$$
%where $B_n(x)$ are Bernoulli polynomials.  
Then we may consider
$$c_k(x)=\sum_{j=0}^k {k \choose j} {{2(2j+2)!} \over {(2\pi)^{2j+2}}} {1 \over
{B_{2j+2}(x)}},  \eqno(49)$$
such that $B_n'(x)=nB_{n-1}(x)$.  If $f(x)=1/B_{2j+2}(x)$, then we have
$${d \over {dx}}f(x)=-f(x)(2j+2){{B_{2j+1}(x)} \over {B_{2j+2}(x)}},  \eqno(50)$$
and Lemma 1 of Ref. \cite{coffeyharm06} applies for the higher order derivatives
of $f(x)$ in terms of (exponential) complete Bell polynomials.  When evaluated
at $x=0$, $B_{2j+1}(0)=0$ unless $j=0$ when $B_1=-1/2$, and Ref. \cite{schimming} describes the Bell polynomials when the odd-indexed variables are set to zero.
%can get some RR for c_k? ...

\medskip
\centerline{\bf Connection with the $\eta_j$ constants}
\medskip

We first mention in passing the following that recovers a result of \cite{baez1},
but in a different way.  We have
{\newline \bf Lemma 3} 
$$\lim_{k \to \infty} P_k(s)(k+1)^s = {1 \over {\Gamma(1-s)}}.  \eqno(51)$$
%on closed and bounded domains of $C$.

{\em Proof}.  We write $P_k(s)=(1-s)_k/k!=\Gamma(k+1-s)/k!\Gamma(1-s)$, apply
the known asymptotic form of $\Gamma(z+a)/\Gamma(z+b)$ (e.g., \cite{nbs}), and
take the limit.

We now introduce the constants $\eta_j$ of the Laurent expansion
$${{\zeta'(s)} \over {\zeta(s)}}=-{1 \over {s-1}}-\sum_{p=0}^\infty \eta_p
(s-1)^p, ~~~~~~~|s-1| < 3,  \eqno(52)$$
with $\eta_0 = -\gamma$.
These coefficients are important in the theory of the function $\ln \zeta(s)$;
hence they are connected with the behaviour of the prime counting function
$\pi(x)$.  The alternating binomial sum $S_2(n) \equiv \sum_{j=1}^k (-1)^j
{k \choose j} |\eta_{j-1}|$ is key in the Li criterion for the RH; its sublinearity
in $n$ would imply the latter conjecture (e.g., \cite{coffeympag}).

%We next invoke Conjecture 1. %, although all we currently require is the much
%weaker statement that the series representation (6) holds unconditionally in a
%neighborhood of $s=1$.  
We apply the identity
$${d \over {ds}}{1 \over {\zeta(s)}}=-{1 \over {\zeta(s)}}{{\zeta'(s)} \over
{\zeta(s)}},  \eqno(53)$$
and evaluate the derivatives as $s \to 1$ from the right.  We obtain
{\bf \newline Proposition 5}.  Put $a_0=\Gamma(k+1/2)/\sqrt{\pi}k!$ and
$$a_q = {1 \over {q!}} {{\Gamma(k+1/2)} \over {\sqrt{\pi} k!}} Y_q
\left[g(s),g'(s),\ldots,g^{(q-1)} (s)\right]_{s=1}, ~~~~~~q \geq 1,  \eqno(54)$$ 
where the function $g$ and its derivatives are given in Eq. (2).
We then have for $q \geq 1$
$$qa_{q+1} \sum_{k=0}^{q+1}c_k=\sum_{j=1}^q a_j \sum_{k=0}^j c_k \eta_{q-j}.
\eqno(55)$$

{\em Proof}.  We first recall from the representation (6) that
$${1 \over {\zeta(s)}}=\sum_{k=0}^\infty c_k \sum_{q=0}^k {1 \over {q!}} 
\left({d \over {ds}}\right)^q \left.P_k\left({s \over 2}\right)\right|_{s=1}
(s-1)^q  \eqno(56a)$$
$$=\sum_{k=0}^\infty c_k \sum_{q=1}^k a_q (s-1)^q.  \eqno(56b)$$
For Eq. (56a) we have kept in mind that $P_k(s)$ is a polynomial of degree $k$
in $s$.  For Eq. (56b) we have used Lemma 1 together with Eq. (5a).  We then
reorder the double summation, obtaining
$${1 \over {\zeta(s)}}=\sum_{q=1}^\infty a_q \sum_{k=0}^q c_k (s-1)^q, \eqno(57a)$$
and
$${d \over {ds}}{1 \over {\zeta(s)}}=\sum_{q=0}^\infty (q+1)a_{q+1} \sum_{k=0}^{q+1}
c_k (s-1)^q. \eqno(57b)$$ 
We then apply identity (53), a form of which is
$$\sum_{k=0}^\infty c_k {d \over {ds}}P_k\left({s \over 2}\right)=
\sum_{k=0}^\infty c_k P_k\left({s \over 2}\right)\left[{1 \over {s-1}}+
\sum_{p=0}^\infty \eta_p (s-1)^p \right].  \eqno(58)$$
We carry out the necessary multiplication of series on the right side of Eq. (58)
and reorder the second term.  We then equate coefficients of like powers of
$s-1$ on both sides and Eq. (55) follows.

%\pagebreak
\medskip
\centerline{\bf Other summatory relations}
\medskip

The authors of the extremely recent Ref. \cite{cislo} derived the identity
$$\sum_{k=0}^\infty c_k s^k = {1 \over {1-s}}\sum_{k=0}^\infty \left ({-s \over
{1-s}}\right)^k {1 \over {\zeta(2k+2)}}, ~~~~~~-1 \leq \mbox{Re} ~s < 1/2  
\eqno(59)$$
and noted the value $\sum_{k=0}^\infty (-1)^k c_k =\sum_{k=1}^\infty 2^{-k}
/\zeta(2k) \simeq 0.7825279853$.

We first illustrate that Eq. (59) can provide the basis of a family of summatory
relations and have
{\newline \bf Proposition 6}.  For $-1 \leq \mbox{Re} ~t \leq 1/2$ we have
$$c_0\ln(1-t)+\sum_{k=1}^\infty {c_k \over k}t^k=\sum_{k=1}^\infty {{(-1)^k} 
\over {k\zeta(2k+2)}} \left({t \over {1-t}}\right )^k,  \eqno(60)$$
giving
{\newline \bf Corollary 5}.
$$c_0\ln(2/3)+\sum_{k=1}^\infty {c_k \over k}{1 \over 3^k}=
\sum_{k=1}^\infty {{(-1)^k} \over {k\zeta(2k+2)}} {1 \over 2^k}
\simeq -0.369410468, \eqno(61a)$$
$$c_0\ln 2+\sum_{k=1}^\infty (-1)^k {c_k \over k}=\sum_{k=1}^\infty {1 \over {k\zeta(2k+2)}} {1 \over 2^k} \simeq 0.65279901499, \eqno(61b)$$
$$-c_0\ln 2+\sum_{k=1}^\infty {c_k \over k}{1 \over 2^k}=\sum_{k=1}^\infty 
{{(-1)^k} \over {k\zeta(2k+2)}} \simeq -0.624463294, \eqno(61c)$$
and
$$\sum_{k=1}^\infty {c_k \over k}\left[(-1)^k+{1 \over 2^k}\right]
=\sum_{k=1}^\infty {1 \over {k\zeta(2k+2)}}\left[(-1)^k+{1 \over 2^k}\right]
\simeq 0.0283357. \eqno(61d)$$

{\em Proof of Proposition 6}.  We write Eq. (59) in the form
$${{c_0 s} \over {s-1}}+\sum_{k=1}^\infty c_k s^k={1 \over {1-s}}
\sum_{k=1}^\infty \left({{-s} \over {1-s}}\right)^k {1 \over {\zeta(2k+2)}}.
\eqno(62)$$
We then divide both sides by $s$, integrate on $s$ from $0$ to $t$, and
Eq. (60) obtains.

Special cases of $t$ in Eq. (60) yield Corollary 5.  

Equations (59) and (60)
may represent the only so far known series associated with the zeta function
where reciprocal zeta values at integer argument occur in the summand.
Equation (59) is extended by 
{\newline \bf Proposition 7}.  For $b >1$ and Re $a>0$ we have
$$\sum_{k=0}^\infty c_k(b,a) s^k = {1 \over {1-s}}\sum_{k=0}^\infty \left (
{-s \over {1-s}}\right)^k {1 \over {\zeta(bk+b,a)}}, ~~~~~~-1 \leq \mbox{Re} ~s < 1/2, 
\eqno(63)$$
where $c_k(b,a)$ is defined in Eq. (19).  In particular, we have
{\newline \bf Corollary 6}.
$$\sum_{k=0}^\infty (-1)^k c_k(b,a) =\sum_{k=1}^\infty {1 \over 2^k}
{1 \over {\zeta(bk,a)}}.  \eqno(64)$$

{\em Proof}.  We use the definition (19), reorder a double sum, and apply the
binomial expansion:
$$\sum_{k=0}^\infty c_k(b,a) s^k=\sum_{k=0}^\infty s^k \sum_{j=0}^k (-1)^j {k 
\choose j} {1 \over {\zeta(bj+b,a)}}$$
$$=\sum_{j=0}^\infty {{(-1)^j} \over {\zeta(bj+b,a)}} \sum_{k=j}^\infty s^k
{k \choose j}$$
$$=\sum_{j=0}^\infty {{(-1)^j s^j} \over {\zeta(bj+b,a)}} \sum_{k=0}^\infty s^k
{{k+j} \choose j}$$
$$=\sum_{j=0}^\infty {{(-1)^j s^j} \over {\zeta(bj+b,a)}} {1 \over {(1-s)^{j+1}}}.
\eqno(65)$$
The alternating sum (64) obtains at $s=-1$.

%A final remark could have the $c_k$ written with the Euler product of $\zeta(s)$ ..
Similarly, Proposition 6 and Corollary 5 may be extended to include the values $c_k(b,a)$.  We have
{\newline \bf Proposition 8}.  For $-1 \leq \mbox{Re} ~t \leq 1/2$,  
$b >1$, and Re $a >0$ there holds
$$c_0(b,a)\ln(1-t)+\sum_{k=1}^\infty {{c_k(b,a)} \over k}t^k=\sum_{k=1}^\infty {{(-1)^k} \over {k\zeta(bk+b,a)}} \left({t \over {1-t}}\right )^k .  \eqno(66)$$ 
We omit the proof.

\medskip
%\pagebreak
\centerline{\bf Third extension of the constants $c_k$}
\medskip

Let $\chi$ be a Dirichlet character mod $k$ and $L(s,\chi)$ the corresponding
Dirichlet $L$-function (e.g., \cite{iwaniec})
$$L(s,\chi)=\sum_{n=1}^\infty {{\chi(n)} \over n^s}, ~~~~~~\mbox{Re} ~s >1.
\eqno(67)$$
We recall that
$${1 \over {L(s,\chi)}}=\sum_{n=1}^\infty {{\chi(n)\mu(n)} \over n^s}, 
~~~~~~\mbox{Re} ~s >1. \eqno(68)$$
We have
{\newline \bf Proposition 9}.  For $b >1$ there holds
$${1 \over {L(s,\chi)}}=\sum_{k=0}^\infty c_k(b,\chi) P_k(s/b), ~~~~~~\mbox{Re}
~s > 1, \eqno(69)$$
where 
$$c_k(b,\chi) \equiv \sum_{j=0}^k (-1)^j {k \choose j} {1 \over {L(bj+b,\chi)}}.
\eqno(70)$$

{\em Proof}.  We proceed as in Ref. \cite{baez1}.  By simply noting that
$|\chi(n)| \leq 1$ the estimates given there justify the interchange of 
infinite summations.  

\medskip
\centerline{\bf Final remarks}
\medskip  

The representation of Proposition 9 is expected to extend to automorphic $L$ functions.  Accordingly we expect a criterion on the rate of growth of 
$|c_k(b,\chi)|$ and its generalization to be equivalent to the extended and generalized Riemann hypothesis, respectively.  

Based upon a special case of Proposition 3 (or 4) there is an extended
Ma\'{s}lanka type representation of Dirichlet $L$ functions.  This follows
since Dirichlet $L$ functions may be written as a combination of Hurwitz 
zeta functions.

The analog of Stieltjes constants and the constants $\eta_j$ exist for
Dirichlet $L$ functions (e.g., Appendix E of Ref. \cite{coffeympag}) and our
method of Proposition 1 or 5 would equally well apply for relating them to $c_k(b,\chi)$.  %cf. p. 172, Ex. 9 of Jameson's book for an
%analog of Stieltjes constants for Dirichlet L funcs.
%also cf. some of the material in one or two of my MPAG paper appendices.

If we insert the Euler product for $\zeta(s)$ into the expression
(1) for $c_k$ we have
$$c_k = \sum_{j=0}^k (-1)^j {k \choose j}\prod_p \left[1-p^{-(2j+2)}\right] ~~~~~~~~~~~~~~~~~~$$
$$=\delta_{0k}-\sum_{j=0}^k (-1)^j {k \choose j}\left[\sum_p p^{-(2j+2)} + \ldots
\right].  \eqno(71)$$
In Eq. (71) the product or sum over $p$ is over all primes and $\delta_{jk}$ is the
Kronecker symbol.  The first sum in brackets on the right side of this equation may
be estimated as $\sum_{p \leq x} p^{-(2j+2)} \sim \mbox{Ei}[-(2j+1) \ln x]$, where 
Ei is the exponential integral.

An approximate expression for $c_k$ for large values of $k$ is given by
\cite{cislo} %Here $c_k \approx R(k)/k$
$$c_k \approx \sum_{n=1}^\infty {{\mu(n)} \over n^2} e^{-k/n^2}.  \eqno(72)$$
We note that alternatively this approximation may be written as a Fourier
transform:
$$c_k \approx {2 \over \pi}\sum_{n=1}^\infty {{\mu(n)} \over n^2}\int_0^\infty
{{\cos k t ~dt} \over {n^2 t^2 + n^{-2}}}.  \eqno(73)$$

\medskip
%\pagebreak
\centerline{\bf Summary}
\medskip

Our results generalize both the B\'{a}ez-Duarte representation of the reciprocal
of the Riemann zeta function \cite{baez1} and the Ma\'{s}lanka representation of
the zeta function itself \cite{maslanka}.  The Ma\'{s}lanka representation has been generalized to the Hurwitz zeta function, the Hurwitz-Lerch zeta function, the
multiple zeta function, and other important special functions.  We have further
extended the B\'{a}ez-Duarte representation of $1/\zeta$ to the representation of
the reciprocal of Dirichlet $L$ functions and it is anticipated that this may be generalized to automorphic $L$ functions.  By way of our generalization of the B\'{a}ez-Duarte representation in terms of Pochhammer
polynomials, we have effectively demonstrated the equivalence of the Riesz
\cite{riesz} and B\'{a}ez-Duarte criteria for the Riemann hypothesis.
We have obtained summatory relations for the coefficients $c_k$ of the 
B\'{a}ez-Duarte criterion for the Riemann hypothesis and related them to 
important constants of analytic number theory.  In describing the relation of
$c_k$ to the Stieltjes and other constants we have made use of the (exponential)
complete Bell polynomials $Y_j$.

%\pagebreak
\bigskip
\centerline{\bf Acknowledgement}   %to consider---
\medskip
%This work was partially supported by Air Force contract number FA8750-04-1-0298.
I thank Prof. L. B\'{a}ez-Duarte for his comments upon reading the manuscript,
in particular in clarifying the statement of Conjecture 1.

\pagebreak
\medskip
\centerline{\bf Appendix:  Interpolating binomial sums from the Fa\`{a} di Bruno formula}
\medskip

%cf. p. 13 of Maslanka (1999)
A significant source of alternating binomial sums is the Fa\`{a} di Bruno formula,
a generalization of the chain rule.  Put $D_z \equiv d/dz$ and $x=x(z)$.  
Then we have
$$D_z^n f(x)=\sum_{k=0}^n D_x^k f(x) {{(-1)^k} \over {k!}} \sum_{j=0}^k
(-1)^j {k \choose j} x^{k-j} D_z^n x^j.  \eqno(A.1)$$
In particular, we have for real $a$
$$D_z^n x^{-a} =a{{a+n} \choose n} \sum_{j=0}^n (-1)^j {n \choose j} {1 \over
{a+j}} x^{-a-j} D_z^n x^j, \eqno(A.2)$$
or equivalently
$$x^a D_z^n x^{-a} =\sum_{j=0}^n {{-a} \choose j}{{n+a} \choose {n-j}} x^{-j}
D_z^n x^j.  \eqno(A.3)$$
In connection with developing alternative representations of analytic functions,
we point out that Eq. (A.3) can be viewed as an immediate consequence of
Lagrange interpolation.  This follows from
$${{-a} \choose j}{{n+a} \choose {n-j}} = \prod_{\stackrel{k=0}{k \neq j}}^n
{{k+a} \over {k-j}}, ~~~~~~~~0 \leq j \leq n.  \eqno(A.4)$$
With the Fa\`{a} di Bruno formula the exponential Bell polynomials again make
an appearance \cite{comtet}.

%\pagebreak
%\medskip
%\centerline{\bf Figure Caption}

%FIG. 1.  ...
%special cases including RK's yields multiplication formula for the digamma
%function ..... TO DO .....

\pagebreak


\begin{thebibliography}{99}
\bibitem{nbs}M. Abramowitz and I. A. Stegun,
{Handbook of Mathematical Functions, Washington, National Bureau of Standards
(1964).}
\bibitem{andrews}G. E. Andrews, R. Askey, and R. Roy,
{Special functions, Cambridge University Press (1999)}.
\bibitem{baez1}L. B\'{a}ez-Duarte,
{A sequential Riesz-like criterion for the Riemann hypothesis, Int. J. Math. Math. 
Sci. {\bf 21}, 3527-3537 (2005).}
\bibitem{baez2}L. B\'{a}ez-Duarte,
{On Ma\'{s}lanka's representation for the Riemann zeta function, math.NT/0307214 (2003).} 
\bibitem{beltra}S. Beltraminelli and D. Merlini,
{The criteria of Riesz, Hardy-Littlewood et al. for the Riemann hypothesis
revisited using similar functions, math.NT/0601138 (2006).}
\bibitem{cislo}J. Cislo and M. Wolf,
{Equivalence of Riesz and B\'{a}ez-Duarte criterion for the Riemann hypothesis,
math.NT/0607782 (2006).}
\bibitem{coffeyharm06}M. W. Coffey,
{A set of identities for a class of alternating binomial sums appearing in
computing applications, to appear in Util. Math. (2007).}
%\bibitem{cofjcam1}M. W. Coffey,
%{On some log-cosine integals related to $\zeta(2)$, $\zeta(3)$, and 
%$\zeta(6)$, to appear in J. Comput. Appl. Math.}
\bibitem{coffeykremin}M. W. Coffey,
{New results on the Stieltjes constants:  Asymptotic and exact evaluation, 
J. Math. Anal. Appl. {\bf 317}, 603-612 (2006); arXiv:math-ph/0506061.}
\bibitem{coffeyprsa}M. W. Coffey,
{New summation relations for the Stieltjes constants, Proc. Royal Soc. A
{\bf 462}, 2563-2573 (2006).}
\bibitem{coffeympag}M. W. Coffey,
{Towards verification of the Riemann hypothesis, Math. Phys., Analysis and
Geometry {\bf 8}, 211-255 (2005).}
\bibitem{comtet}L. Comtet,
{Advanced Combinatorics, D. Reidel (1974).}
%\bibitem{edwards}H. M. Edwards,
%{Riemann's Zeta Function, Academic Press, New York (1974).}
%\bibitem{grad}I. S. Gradshteyn and I. M. Ryzhik,
%{Table of Integrals, Series, and Products, Academic Press, New York (1980).}
\bibitem{ivic}A. Ivi\'{c}, 
{The Riemann Zeta-Function, Wiley (1985).}
\bibitem{iwaniec}H. Iwaniec and E. Kowalski,
{Analytic number theory, American Mathematical Society (2004).}
\bibitem{maslanka}K. Ma\'{s}lanka,
{B\'{a}ez-Duarte criterion for the Riemann hypothesis and Rice's integrals,
math.NT/0603713 v2 (2006).}
\bibitem{riesz}M. Riesz,
{Sur l'hypoth\`{e}se de Riemann, Acta Math. {\bf 40}, 185-190 (1916).}
\bibitem{riordan1}J. Riordan,
{An introduction to combinatorial analysis, Wiley (1958).}
\bibitem{riordan2}J. Riordan,
{Combinatorial identities, Wiley (1968).}
\bibitem{schimming}R. Schimming and W. Strampp,
{Differential polynomial expressions related to the Kadomtsev-Petviashvili and
Korteweg-de Vries hierarchies, J. Math. Phys. {\bf 40}, 2429-2444 (1999).}
\bibitem{vardi}I. Vardi,
{Determinants of Laplacians and multiple gamma functions, SIAM J. Math. Analysis
{\bf 19}, 493-507 (1988).}
\bibitem{wolf}M. Wolf,
{Evidence in favor of the B\'{a}ez-Duarte criterion for the Riemann hypothesis,
math.NT/0605485 (2006).}
%& recall Warren Johnson's AMM (2000) article on The curious history of
%Faa di Bruno's formula ...-
\end{thebibliography}
\end{document}